\documentclass[final]{raa06}
\usepackage{graphicx,times}             
\usepackage{natbib}
\usepackage{amssymb,amsmath}
\bibpunct{(}{)}{;}{a}{}{,}

\newcommand{\cm}{{~\rm cm}}

\newcommand{\km}{{~\rm km}}
\newcommand{\s}{{~\rm s}}

\newcommand{\g}{{~\rm g}}

\newcommand{\erg}{{~\rm erg}}

\newcommand{\MeV}{{~\rm MeV}}

\begin{document}

   \title{Boosting jittering jets by neutrino heating in core collapse supernovae
}

   \volnopage{Vol.0 (20xx) No.0, 000--000}      
   \setcounter{page}{1}          

   \author{Noam Soker
      \inst{1}
   }

   \institute{Department of Physics, Technion, Haifa, 3200003, Israel;  soker@physics.technion.ac.il {\it soker@physics.technion.ac.il}\\
\vs\no
   {\small Received~~20xx month day; accepted~~20xx~~month day}}
\abstract{
I estimate the energy that neutrino heating adds to the outflow that jets induce in the collapsing core material in core collapse supernovae (CCSNe), and find that this energy crudely doubles the energy that the jets deposit into the outer core. I consider the jittering jets explosion mechanism where there are several stochastic jet-launching episodes, each lasting for about 0.01-0.1 seconds. The collapsing core material passes through the stalled shock at about 100 km and then slowly flows onto the proto-neutron star (NS). I assume that the proto-NS launches jittering jets, and that the jets break out from the stalled shock. I examine the boosting process by which the high-pressure gas inside the stalled shock, the gain region material, expands alongside the jets and does work on the material that the jets shock, the cocoon. This work is crudely equal to the energy that the original jets carry. I argue  that the coupling between instabilities, stochastic rotation, magnetic fields, and jittering jets lead to most CCSN explosions. In other cases, the pre-collapse core is rapidly rotating and therefore ordered rotation replaces stochastic rotation and fixed jets replace jittering jets. 
\keywords{core collapse supernovae; stellar jets} }

 \authorrunning{N. Soker}            
   \titlerunning{Boosting jittering jets in CCSNe}  
   
      \maketitle

\section{Introduction} 
\label{sec:intro}

The collapsing inner core of a massive star in a core collapse supernova (CCSN) releases $>{\rm few} \times 10^{53} \erg$ in gravitational energy. The rest of the star explodes and carries a fraction of $\simeq 0.0001-0.1$ of this energy (e.g., \citealt{Hegeretal2003, Janka2012}). Neutrinos carry the rest of the energy, while typically only a small fraction of the released gravitational energy ends in radiation. The very inner part of the core forms a nuclear-density compact object at the center, the proto-neutron star (NS). The formation of the pro-NS stops the collapse of the very inner part of the core and sends a shock wave outward. Because of the ram pressure of the collapsing core material the shock stalls at a radius of about $R_{\rm s} \simeq 100 \km$. This is the stalled shock. The mass that the proto-NS accretes flows through this stalled shock. 

Recent theoretical studies consider two mechanisms to channel a small fraction of the gravitational energy to the exploding outer core and envelope, in case an envelope exists. According to the delayed neutrino mechanism \citep{BetheWilson1985} neutrinos heat the material in the post-shock zone behind the stalled shock, the gain region, and after some delay the heating revives the stalled shock in a non-spherical manner (e.g., \citealt{Nordhausetal2012, CouchOtt2013, Bruennetal2016, Jankaetal2016R, OConnorCouch2018, Mulleretal2019Jittering, BurrowsVartanyan2021, Fujibayashietal2021, Bocciolietal2022, Nakamuraetal2022}).
According to the jittering jets explosion mechanism \citep{Soker2010} the proto-NS, or later the newly born NS or a black hole (BH) if it is formed, launches jets that deposit sufficient energy to the collapsing-core material outside the stalled shock and explode the star (e.g., \citealt{PapishSoker2011, PapishSoker2014Planar, GilkisSoker2015, Quataertetal2019, Soker2020RAA, AntoniQuataert2022, ShishkinSoker2022, Soker2022a}).

 During the explosion there might be several to few tens of jet-launching episodes. Some typical physical values are as follows \citep{PapishSoker2014a}. The jets carry a total energy of $\simeq 10^{51} \erg$ with a typical velocity of $\simeq 10^5 \km \s^{-1}$ and total explosion time of $\simeq 1 \sec$. Each jet-launching episode lasts for $\simeq 0.01-0.1 \sec$ and carries a mass of $\approx 10^{-3} M_\odot$. The accretion disk that launches the jets is ten times as massive, so each accretion disk mass at a jet-launching episode is $\approx 10^{-2} M_\odot$. In total, during the entire explosion process a mass of $\approx 0.1 M_\odot$ is accreted through an intermittent accretion disk. 

The source of the stochastic angular momentum of the mass that is accreted through the intermittent accretion disk is the stochastic convection motion in the pre-collapse core. The convective cells in the silicon or oxygen burning zones serve as perturbation seeds that instabilities above the newly born NS further amplify to supply stochastic angular momentum with large enough amplitudes to form intermittent accretion disks (e.g., \citealt{ShishkinSoker2022}).  An important property of the jittering jets explosion mechanism is that the jets of early episodes do not disturb these perturbations for later jet-launching episodes. The jets do influence somewhat the directions of the later jittering jets \citep{PapishSoker2014Planar}. The source of magnetic fields that are required to launch jets is also in the pre-collapse core, from the convective zones as well as the radiative zone above the iron core \citep{Perersetal2019}. The amplification of the seed angular momentum perturbations, the role of the magnetic field in influencing the stochastic angular momentum (by angular momentum transfer), and the interaction of neutrino-driven convection plumes with the accretion disks are open questions to be determined by future studies of the jittering jets explosion mechanism. 

The jittering jets explosion mechanism differs from many studies of jet-driven explosions that assume rapidly rotating pre-collapse cores (e.g., \citealt{Khokhlovetal1999, Aloyetal2000, MacFadyenetal2001, Maedaetal2012, LopezCamaraetal2013, BrombergTchekhovskoy2016,  Nishimuraetal2017, WangWangDai2019RAA, Grimmettetal2021, Perleyetal2021}) in the following properties. (1) The jittering jets operate in a negative feedback mechanism (see review by \citealt{Soker2016Rev}). This explains why typical explosion energies are several times the binding energy of the ejected mass. (2) According to the jittering jets explosion mechanism most (or even all) CCSNe are driven by jets, even CCSNe of non-rotating cores. In cases of slowly rotating (or non-rotating) pre-collapse cores, the convective motion in the pre-collapse core (e.g., \citealt{GilkisSoker2014, GilkisSoker2016, ShishkinSoker2021}) or envelope (e.g., \citealt{Quataertetal2019}) allows the formation of intermittent/stochastic accretion disks or belts that launch jittering jets. Namely, the accretion process through intermittent accretion disks/belts leads to jet-launching episodes where the direction of the symmetry axis of the two opposite jets changes from one episode to the next. 

One result of these differences is that there are no failed CCSNe in the jittering jets explosion mechanism framework (e.g., \citealt{Gilkisetal2016Super, Soker2017RAA, AntoniQuataert2022}). This claim has received strong support with the new observational finding by \cite{ByrneFraser2022}. According to the jittering jets explosion mechanism all massive stars explode, even when the explosion forms a BH.
In the jittering jets explosion mechanism the explosion leads to BH formation in the case of a rapidly rotating pre-collapse core. The jets that the newly born NS launches maintain a fixed axis along the large angular momentum axis, and therefore the jets eject only a small fraction of the stellar mass along the polar directions. The rest of the stellar mass collapses to form a BH. The jets lead to a very energetic CCSN, i.e., some of the most energetic CCSN explosions are those that form BHs (e.g., \citealt{Gilkisetal2016Super, Soker2017RAA}). On the other hand, according to the delayed neutrino mechanism it is possible that a massive star does not explode, but rather most of the mass collapses to form a BH accompanied by a faint transient event (e.g., \citealt{Nadezhin1980, LovegroveWoosley2013}). 

There is no need to form a thin accretion disk that is supported against gravity solely by the centrifugal force to launch jets, although this is the case in many astrophysical objects. Consider a case where the specific angular momentum is smaller than the critical value to support a Keplerian motion around the newly born NS (proto-NS), but the accretion flow does form low-density funnels along the two opposite polar directions (along the angular momentum axis). \cite{SchreierSoker2016} argued that this accretion belt geometry allows the launching of jets along the polar funnels. Critical to the launching of jets from accretion belts in CCSNe is the presence of very strong magnetic fields (e.g., \citealt{SchreierSoker2016, Soker2018arXiv, Soker2019RAA, Soker2020RAA}). 
This claim, that the compact object at the center of a CCSN can launch jets even when the core does not rotate, but only if there are strong magnetic fields and funnels along the polar directions, has received indirect support from recent three-dimensional magneto-hydrodynamical simulations of a BH moving through a uniform magnetized medium and accreting mass. In these simulations, \cite{Kaazetal2022} find that a BH can launch strong jets despite that the initial angular momentum of the accreted gas is zero, as long as the magnetic fields are sufficiently strong. The ordered magnetic fields that they use form the funnels along the symmetry axis. 

In some earlier papers (e.g., \citealt{Soker2019arXiv}) I mentioned the possibility that there is a mutual influence between stochastic angular momentum accretion and neutrino heating, and that the combined operation of jets and neutrino heating power CCSNe. In the present paper I study in more detail the way by which neutrino heating in the gain region can boost the outflow that the jets induce. I recall that each jet launching episode is expected to last for $<0.1 \s$. Therefore, although the jets of the first pair of jets  break out from the stalled shock, there is no time to set an explosion via neutrino heating alone because the delayed neutrino mechanism requires a much longer time to set an explosion (e.g., \citealt{Bolligetal2021}). For that, I expect that there will be at least several jet-launching episodes before the end of the explosion process.   
I present the relevant parameters of the gain region and of the jets in section \ref{sec:RelevantParameters}, and the processes by which the neutrino heating boost the outflow in section \ref{sec:Boosting}.  In section \ref{sec:Assumptions} I discuss some of the assumptions that I make in this study.  
I summarize in section \ref{sec:Summary}.  
 
\section{Relevant parameters} 
\label{sec:RelevantParameters}

\subsection{The ambient medium} 
\label{subsec:Ambient}

In scaling the quantities for the ambient medium, namely, the medium into which the jets expand, I make use of the detailed study by \cite{Janka2001}. I chose specific profiles to demonstrate the boosting processes, but the results are not sensitive to these specific profiles. 

I consider the time in the collapse when the stalled shock is more or less static at $R_{\rm s}=100 \km$ and the gain radius is at $R_{\rm g}=50 \km$. The gain radius is defined to be such that in the zone $R_{\rm g} < r < R_{\rm s}$ neutrino heating overcomes neutrino cooling. This zone is made of free nucleons and alpha particles.    
For the density profile I take (equation 63 scaled with figure 3 from \citealt{Janka2001}) 
\begin{equation}
\rho_{\rm g} \simeq 2 \times 10^9
\left( \frac{r}{R_{\rm s}} \right)^{-3}
\g \cm^{-3} ; \quad 50 \km < r< 100 \km.
\label{eq:RhoG}
\end{equation}
Note that I consider a spherical and static stalled shock front and ignore here the standing accretion shock instability (SASI), which I return to in section \ref{subsec:SASI}. 

For an adiabatic index of $\gamma=4/3$ the pre-shock gas density is $\rho_{\rm p}=\rho_{\rm g} (R_{\rm s})/7$. As photodissociation of nuclei increases this density ratio (e.g., \citealt{Thompson2000}), I take this ratio to be 8. For a mass of $M_{\rm s} = 1.4 M_\odot$ inside the stalled shock and a free fall velocity of the pre-shock gas, $v_{\rm ff,s}=6.1\times 10^4 \km \s^{-1}$, the accretion rate of the collapsing core at the shock is $\dot M_{\rm c} \simeq 1 M_\odot \s^{-1}$. I do note that the collapsing velocity is somewhat smaller than the free fall velocity (e.g., \citealt{Janka2001}), but like \cite{Thompson2000} I take it to be the free fall velocity. The post-shock pressure is about equal to the pre-shock ram pressure, and for the pressure profile in the gain region I take (equation 63 from \citealt{Janka2001}) 
\begin{equation}
P_{\rm g} \simeq 10^{28}
\left( \frac{r}{R_{\rm s}} \right)^{-4}
\erg \cm^{-3} ; \quad 50 \km < r< 100 \km.
\label{eq:PressG}
\end{equation}
Photons and electron-positron pairs dominate the pressure, and so the temperature is 
\begin{equation}
k T_{\rm g} \simeq 3
\left( \frac{r}{R_{\rm s}} \right)^{-1}
\MeV ; \quad 50 \km < r< 100 \km.
\label{eq:TG}
\end{equation}
The internal energy in the gain region is $E_{\rm th,g} \simeq \int 3 P_{\rm g} dV \simeq 4 \times 10^{50} \erg$. In the present study I examine the possibility that a fraction of this energy boosts the jets. This energy is not the total energy in the gain region during the entire explosion time, but rather the energy at a given time. The gain region loses energy but neutrino heating and the in-flowing core material replenish the energy. Therefore, the total energy that the gain region can add to the outflow during the entire explosion process is $>E_{\rm th,g}$. 

\subsection{The jets} 
\label{subsec:Jets}

I do not study the formation of jets (see section \ref{sec:intro}  and section \ref{subsec:Launching}).  I assume that the proto-NS (or newly born NS) launches the jets inside the gain region, i.e., at $r < R_{\rm g} \simeq 50 \km$. The inner boundary of the launching zone is the proto-NS. The neutrinosphere is outside the proto-NS at very early times, moving into the proto-NS at late times (e.g., \citealt{Jankaetal2007}). As the jets are launched by the operation of magnetic fields, the location of the neutrinosphere is not of large significance.  The jets might start with a mass outflow rate that is $\simeq 0.1-0.2$ times the mass accretion rate, and have a terminal velocity larger than the escape velocity from the proto-NS, which is somewhat larger than $10^5 \km \s^{-1}$. Like \cite{PapishSoker2011} I take the velocity of the gas in the jets before it is shocked to be constant at $v_{\rm j} =10^5 \km \s^{-1}$, and its mass outflow rate to be $\dot M_{\rm 2j} \simeq 0.1 \dot M_{\rm c}$. The deceleration by the gravity of the central mass is significant at short distances of $r \la 100 \km$. I am here interested in the interaction mainly near and outside the stalled shock, i.e., at  $r \ga R_{\rm s}$, and so I take the above value of jets velocity at $R_{\rm s}$, and for the accuracy of the present study I can neglect gravitational deceleration at $r>R_{\rm s}$. The solid angle that the two opposite jets cover is 
\begin{equation}
\Omega_{\rm 2j} = 4 \pi \delta,
\label{eq:OmegaJets}
\end{equation}
where $\delta$ can vary as the jets expand, i.e., $\delta(r,t)$. Like \cite{PapishSoker2011} I scale with $\delta=0.01$ that corresponds to a half-opening angle of $\alpha_{\rm j}=8^\circ$. 
The density in the jets is  
\begin{equation}
\begin{split}
\rho_{\rm j}= 1.6 \times 10^9  
\left( \frac{\dot M_{\rm 2j}}{0.1 M_\odot \s^{-1}} \right)
\left( \frac{v_{\rm j}}{10^5 \km \s^{-1}}\right)^{-1}   
\\ \times \left( \frac{\delta}{0.01} \right)^{-1}
\left( \frac{r}{100 \km} \right)^{-2}
\g \cm^{-3} ; \qquad r > 50 \km  .
\label{eq:RhoJets}
\end{split}
\end{equation}
Each jet-launching episodes lasts for $\tau_{\rm j} \simeq 0.01-0.1 \s$ and carry and energy of $E_{\rm 2j} \approx {\rm few} \times 10^{49} - {\rm few} \times 10^{50} \erg$. Note that the jets get their initial energy directly from the accretion energy of the accretion disk, and not from the gain region. The energy from the gain region boosts the outflow that the jets induce.  

A possible case for the present setting is one where there are $N_{\rm launch} = 4$, or few more, jet-launching episodes as I claimed for the supernova remnant SNR~0540-69.3 \citep{Soker2022a}, each lasting $\tau_{\rm j} \simeq 0.03 \s$ or somewhat less. The total energy of the four jet-launching episodes is $1.2 \times 10^{51} \erg$, which after adding the contribution from neutrino heating that I study here and removing the binding energy of the ejecta gives a typical CCSN explosion energy, i.e., $\simeq 10^{51} \erg$. By explosion energy I refer to the kinetic energy of the ejecta, which dominates the explosion energy, plus the radiated energy.

\begin{figure*}
\begin{center}
\includegraphics[trim=28.5cm 8.2cm 29.0cm 2.0cm,scale=0.86]{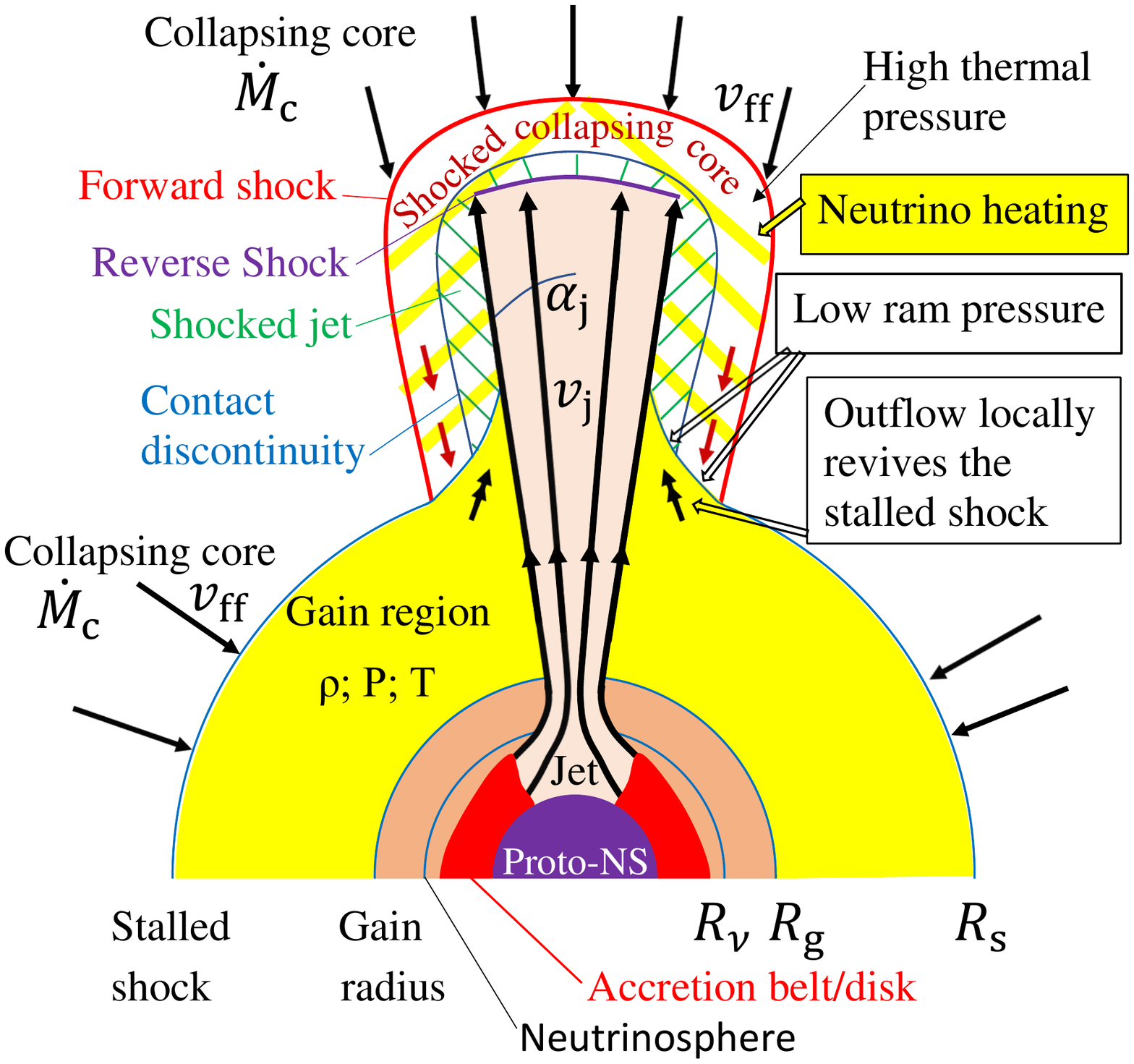} 
\caption{A schematic drawing (not to scale) of the flow structure of one of the two opposite jets just after it breaks out from the gain region. The effects inside boxes that point with double-line arrows are those involved in boosting the outflow that the jets induce. The yellow-hatched region is the cocoon, which includes the post-shock material of the jets and of the collapsing core.  In the figure the neutrinosphere is outside the proto-NS, which is true only at the very beginning of the process, as in a short time it moves into the proto-NS (e.g.,  \citealt{Jankaetal2007}). }
\label{Fig:Schematic}
\end{center}
\end{figure*}

The outcome for the parameters that I use here is that the density in the jets is similar to the density in the gain region. Since the jets' receive the same amount of neutrino flux from the center, the temperature and pressure inside the pre-shock jets is similar to that in the gain region.
Actually, if the thermal pressure in the ambient gas is larger than that inside the jet the ambient medium compresses the jet, while if the ambient thermal pressure is smaller then the cross section of the jet increases. Therefore, the thermal pressures inside the jets and in the ambient gas are about equal. As the pressure close to the proto-NS is much larger than that in the outer gain region (e.g., \citealt{Janka2001}), the jets are expected to be collimated. I present the schematic flow structure in Fig. \ref{Fig:Schematic}.
     
Momentum balance at the head of a jet implies that within the gain region the jet's head velocity is (e.g., \citealt{PapishSoker2011})
\begin{equation}
v_{\rm h,g} \simeq 0.5 v_{\rm j} 
\left( \frac{2}{1+\sqrt{\rho_{\rm g}/{\rho_{\rm j}}}} \right).
\label{eq:Vhg}
\end{equation}
The jet's head crosses the gain region within $\Delta t_{\rm h,g} \simeq 0.001 \s \la 0.1 \tau_{\rm j}$.

As the jets propagate out from the stalled shock they encounter the collapsing core material. 
The ram pressure of the collapsing gas in the frame of the stationary stalled shock $P_{\rm ram,p}$ just before it hits the stalled shock is about the post-shock pressure as given by equation (\ref{eq:PressG}), and varies as $P_{\rm ram,p} \simeq P(R_{\rm s}) (r/R_{\rm s})^{-5/2}$ (e.g.  \citealt{BlondinMezzacappa2006}).  
The ratio of the jets' ram pressure to that of the collapsing gas in the frame of the stationary stalled shock is
\begin{equation}
\begin{split}
\frac{\rho_{\rm j} v^2_{\rm j}}{P_{\rm ram,p}} \simeq 16 
\left( \frac{\dot M_{\rm 2j}}{0.1 \dot M_{\rm c}} \right)
\left( \frac{v_{\rm j}}{10^5 \km \s^{-1}}\right) \\ \times 
\left( \frac{\delta}{0.01} \right)^{-1}
\left( \frac{r}{100 \km} \right)^{1/2}
; \quad r > R_{\rm s} = 100 \km
\label{eq:eq:RamPs}
\end{split}
\end{equation}
and I take $R_{\rm s}=100 \km$. 
Note that in this approximation I neglect the time variation of the accretion rate (the jets were launched by earlier accreted gas). 
 
To find the velocity of the head of a jet $v_{\rm h}$ as the jet propagates through the collapsing core material we need to equate the ram pressures in the frame of the jet's head 
\begin{equation}
\rho_{\rm c} (v_{\rm h} + v_{\rm ff})^2 \simeq 
\rho_{\rm j} (v_{\rm j} - v_{\rm h})^2,
\label{eq:RamPressurs1}
\end{equation}
where I will take for the collapsing core velocity the free fall velocity that I define positively, $v_{\rm ff} > 0$, and the density of the collapsing material is $\rho_{\rm c} = \dot M_{\rm c}/ (4 \pi r^2 v_{\rm ff})$.
The density of the pre-shock material inside the jet is given equation (\ref{eq:RhoJets}), and I define 
\begin{equation}
\beta \equiv \frac {\dot M_{\rm 2j}}{\delta \dot M_{\rm c}} = 10
\left( \frac{\dot M_{\rm 2j}}{0.1 \dot M_{\rm c}} \right)
\left( \frac{\delta}{0.01} \right)^{-1}.
\label{eq:beta}
\end{equation}
Equation (\ref{eq:RamPressurs1}) is a quadratic equation for $v_{\rm h}$ that reads 
\begin{equation}
\begin{split}
\left( \beta v_{\rm ff} - v_{\rm j}\right) v^2_{\rm h} 
- 2 v_{\rm j} v_{\rm ff} \left(1+ \beta\right)v_{\rm h} 
+ v_{\rm j} v_{\rm ff} \left( \beta v_{\rm j} - v_{\rm ff}  \right)=0.
\end{split}
\label{eq:RamPressurs2}
\end{equation}

Just outside the stalled shock at $R_{\rm s} =100 \km$ the velocity of the jet's head is $v_{\rm h}=0.54 v_{\rm j}$, where I substituted the parameters I use here $v_{\rm j}=10^5 \km \s^{-1}$, $\beta=10$, and $v_{\rm ff}(R_{\rm s})=6.1 \times 10^4 \km \s^{-1}$. \cite{PapishSoker2011} use a different approach to calculate the jet's head velocity, i.e., they take the ambient density from numerical simulations. Using the same parameters as I do here, their equation (6) gives a similar value of $v_{\rm h} \simeq 0.54 v_{\rm j}$. 
At $r=1600 \km$ where $v_{\rm ff}= 1.5 \times 10^4\km \s^{-1}$ the solution of equation (\ref{eq:RamPressurs2}) gives a similar value of $v_{\rm h}=0.48 v_{\rm j}$. Namely, the velocity of the jet's head does not change much near and outside the radius of the stalled shock, and for the parameters that I use here it is 
\begin{equation}
v_{\rm h} \simeq 0.5 v_{\rm j}. 
\label{eq:VhOut}
\end{equation}

\begin{figure}
\begin{center}
\includegraphics[trim=28.5cm 3.2cm 30.0cm 2.0cm,scale=0.58]{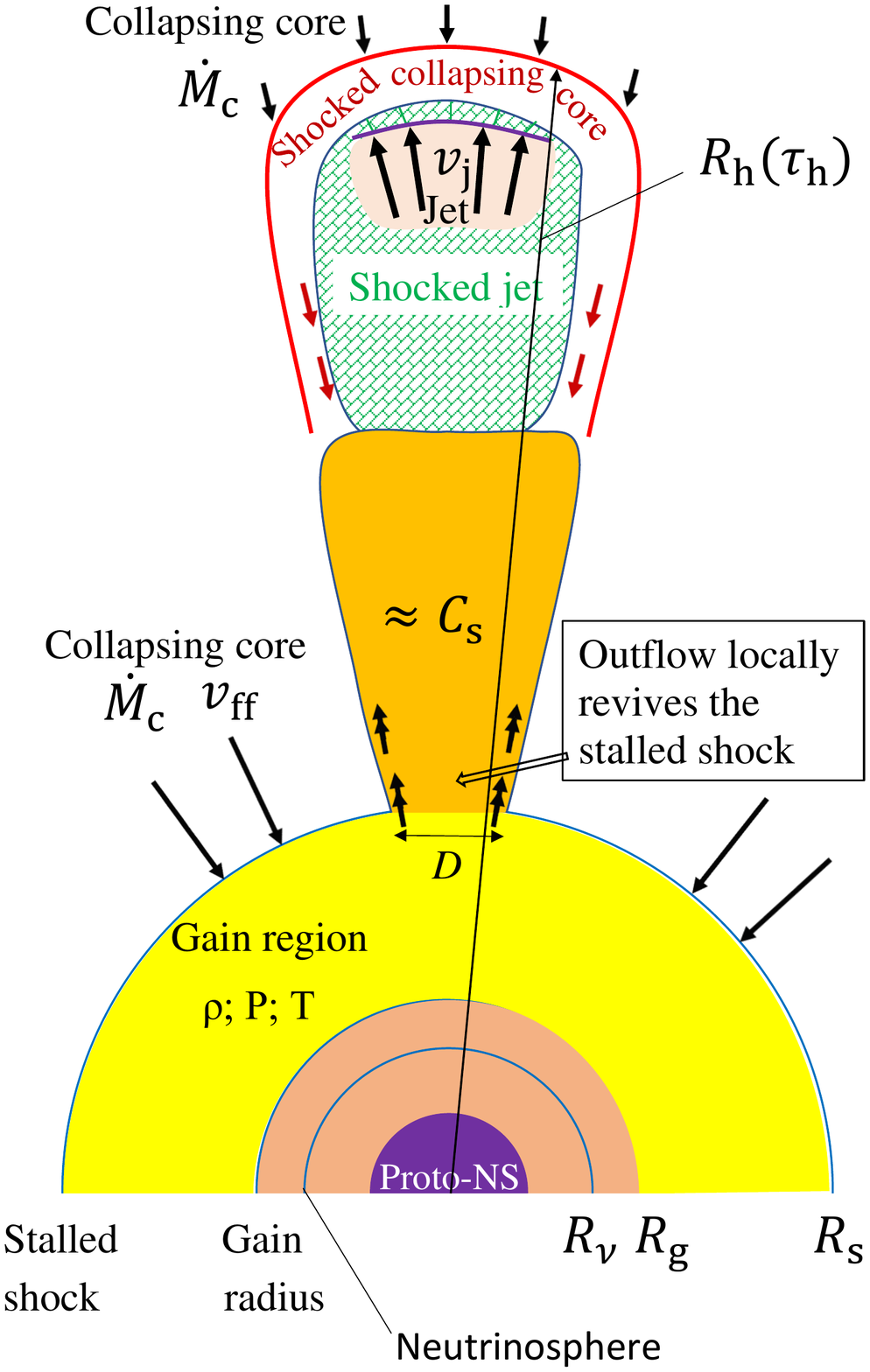} 
\caption{A schematic drawing (not to scale) of the flow structure of one of the two opposite jets just as the jet ceases to feed the head, at $\tau_{\rm h}$ as given by equation (\ref{eq:Tauhead}).  Note that at late times, unless accretion rate increases a lot, the neutrinosphere is inside the proto-NS.  }
\label{Fig:Schematic2}
\end{center}
\end{figure}

For a jet-launching episode that lasts $\tau_{\rm j} \simeq 0.03 \s$ the jet reaches
a distance $\simeq 1500 \km$ by the end of the episode. However, the head of the jet will `know' about this at a much later time. 
For example, if the jet's head proceeds at half the velocity of the material in the jets it will reach a distance of $\simeq 3000 \km$ before the supply of fresh jet's material ceases. This distance is similar to simulations of the jittering jets \citep{PapishSoker2014a, PapishSoker2014Planar}. For a constant jet's head velocity the interaction will continue until time
\begin{equation}
\tau_{\rm h} \simeq 2 \tau_{\rm j} \frac{v_{\rm j}}{2 \left( v_{\rm j}- v_{\rm h} \right) },
\label{eq:Tauhead}
\end{equation}
from the beginning of the jet-launching episode, when the jet's head reaches a distance of 
\begin{equation}
R_{\rm h}(\tau_{\rm h}) \simeq 3000 
\left( \frac{\tau_{\rm j}}{0.03 \s} \right) 
\left( \frac{v{\rm j}}{10^5 \km \s^{-1}} \right) 
\frac{v_{\rm h}}{v_{\rm j}- v_{\rm h}} \km. 
\label{eq:Rhead}
\end{equation}
In Fig. \ref{Fig:Schematic2} I schematically (not to scale) draw the flow at the time $\tau_{\rm h}$ when the jet ceases to feed the head.

The interaction of the jets with the collapsing core is more complicated because neutrino heating and cooling take place in the pre-shock and post-shock media, that of the jets and that of the collapsing core. The post-shock regions of the two media is the `cocoon'. The shock wave that runs into the collapsing core (forward shock) has a size larger than the jet's radius $r \sin \alpha_{\rm j}$.    

In the rest frame of the jet's head the collapsing core is shocked at a velocity that is close to about twice as large as its free fall velocity at the stalled shock. This implies that the pressure of the post-shock material of the collapsing core, the `cocoon', is about three times as large as that just behind the stalled shock. Therefore, the cocoon maintains the collimation of the jet, or might even compress it somewhat.  

\section{The boosting process} 
\label{sec:Boosting}

In the jittering jets explosion mechanism there are in total several to few tens of jets-launching episodes, each lasting $\tau_{\rm j} \simeq 0.01-0.1 \s$, or even up to few$\times 0.1 \s$. In each episode the accretion belt/disk launches two opposite jets. The jets' axis directions change by a large angle from one episode to the next. 
Although in Fig. \ref{Fig:Schematic}, that represents one jet in one episode, I draw a straight jet, the jets direction might continuously change (precesses) by a small angle even within a jet-launching episode. 

The interaction of each of the two opposite jets with the collapsing core material shocks both the jet's material, in the reverse shock, and the collapsing core material, in the forward shock. These two post-shock zones, separated by a contact
discontinuity, are referred to as the ‘cocoon’ (yellow-hatched region in Fig. \ref{Fig:Schematic}). 
There are two processes by which neutrino heating boosts the outflow of the `cocoon' along the sides of the jets.  

\subsection{Heating the cocoon} 
\label{subsec:HeatingCocoon}

In the direct heating process that I consider first neutrinos directly heat the cocoon. To estimate the importance of this process I estimate the properties of the cocoon. 
Equation (\ref{eq:VhOut}) implies that as the jets expand into the collapsing core, at least for $r \la 2000 \km$, both the jets' material and the collapsing core that the jets interact with pass through strong shocks, e.g., shock velocities of $\simeq 5 \times 10^4 - 10^5 \km \s^{-1}$. This is similar to the velocity by which the collapsing core hits the stalled shock away from the jets. 
Therefore, the post-shock jets and collapsing core zones, the `cocoon' (see Fig. \ref{Fig:Schematic}), are nearly fully composed of free nucleons. This in turns implies an efficient neutrino heating, similar to that in the gain region. 

Neutrino heating depends on the neutrino flux $L_{\nu} / 4 \pi r^2$ and neutrino energy, where the relevant neutrinos are the electron-neutrino and the anti-electron-neutrino. 
As typical neutrino luminosity and average energy in the first second of collapse for a model with an initial mass of $M_{\rm ZAMS} = 15 M_\odot$ I take $L_\nu = 2 \times 10^{52} \erg \s^{-1}$ and $\bar{e}_\nu =12 \MeV$, respectively, and note that different models might have different values (e.g, \citealt{Mulleretal2012, Glasetal2019, Matsumotoetal2022, Nakamuraetal2022}). The neutrino heating rate per nucleon at a distance $r$ from the center is (e.g., \citealt{Janka2001})
\begin{equation}
\begin{split}
q_{\nu} \approx  3000
\left( \frac{r}{100 \km} \right)^{-2}
\left( \frac{L_{\rm \nu}}{2 \times 10^{52} \erg \s^{-1}} \right)
\\ \times 
\left( \frac{\bar{e}_\nu }{12 \MeV} \right)^{2}
\MeV \s^{-1}.
\label{eq:Qheat}
\end{split}
\end{equation}

I present here a very crude estimate, since only hydrodynamical simulations of the interaction of the jets with the collapsing core can find the total heating of the cocoon. Consider that the jets shock and drag out collapsing core material of equal mass. This means that each nucleon in the jets drag another nucleon from the collapsing core. Although the interaction region moves out, there must be material between the jet's head and the stalled shock. I consider that this material flows outward at a speed of $v_{\rm cocoon} \approx 0.5 v_{\rm j} \simeq 5 \times 10^4 \km \s^{-1}$. 
To find the total heating of the cocoon per nucleon in the jet I 
integrate over time $dt = dr /v_{\rm cocoon}$ from $r=100 \km$ out and multiply by two, as for each nucleon in the jet there are two in the cocoon. I find that for each original nucleon in the jet direct neutrino heating of the cocoon adds energy of 
\begin{equation}
\begin{split}
\Delta E_{\rm n_j, \nu} <  10
\left( \frac{L_{\rm \nu}}{2 \times 10^{52} \erg \s^{-1}} \right)
\left( \frac{\bar{e}_\nu }{12 \MeV} \right)^{2}
\\ \times 
\left( \frac{v_{\rm cocoon}}{5 \times 10^4 \km \s^{-1}} \right)^{-1}
\MeV. 
\label{eq:DeltaEn}
\end{split}
\end{equation}
  I use the inequality sign above because this calculation overestimates the time the cocoon stays close to the stalled shock and does not include neutrino cooling.  Including the appropriate time and neutrino cooling reduce the energy that equation (\ref{eq:DeltaEn}) would give to less than half the value,  i.e., $\Delta E_{\rm n_j, \nu} < 5 \MeV$ (see also \citealt{Janka2001}). The kinetic energy that each nucleon in the jet carries, which is 
$E_{\rm n_j,k} = 57 \MeV$ for $v_{\rm j} = 10^5 \km \s^{-1}$,  is about an order of magnitude larger. Therefore, direct neutrino heating of the cocoon plays a very small role in boosting the explosion by jets.

The significant conclusion of this subsection is that future numerical simulations of the neutrino boosting of jittering jets, that will be highly resources-demanding, can ignore the direct heating of the cocoon.  

\subsection{Accelerating the cocoon out} 
\label{subsec:AcceleratingCocoon}

With the parameters of equations (\ref{eq:RhoG})-(\ref{eq:TG}) the sound speed of the gain region at $r\la R_{\rm s}$ is 
\begin{equation}
C_{\rm s} (R_{\rm s}) \simeq 2.5 \times 10^4 \km \s^{-1} \simeq 0.5 v_{\rm h}, 
\label{eq:Cs}
\end{equation}
where in the second equality I used equation (\ref{eq:VhOut}).
The collapsing core material along the jet's axis and around is shocked by the jet, and therefore reaches the stalled shock at a slower velocity than that of the undisturbed collapsing core material. As a result of that the ram pressure on the stalled shock around the jet, namely, where the cocoon touches the stalled shock (see Fig. \ref{Fig:Schematic}), is much lower than that of the undisturbed collapsing core material. This leads the high-pressure gas just behind the stalled shock to stream into the cocoon (the arrows with a double-triangle head). This adds energy to the cocoon.
In other words, the material in the gain region locally revives the stalled shock (Fig. \ref{Fig:Schematic}) and does work on the cocoon. 

Only full 2D or 3D hydrodynamical simulations can determine the extra work that the gas from the gain region does on the cocoon.
I here only crudely estimate a plausible value. 
Consider that the cocoon cross section on the stalled shock is $b \simeq {\rm few}$ times the cross section of the jet. The combined cross sections of the two opposite cocoons (only one shown in Fig. \ref{Fig:Schematic}) is 
\begin{equation}
A_{\rm cocoon}=b 4 \pi \delta R^2_s. 
\label{eq:Acocoon}
\end{equation}
The rate of work done on the cocoon by the material of the gain region that expands behind the cocoon is 
\begin{equation}
\begin{split}
W_{\rm g} & \approx A_{\rm cocoon} P(R_s) C_{\rm s} 
\simeq 9.4 \times 10^{50}
\left( \frac{b}{3} \right) 
\left( \frac{\delta}{0.01} \right) 
\\ \times &
\left( \frac{R_{\rm s}}{100 \km} \right)^2
\left( \frac{P(R_s)}{10^{28} \erg \cm^{-3}} \right)
\\ \times &
\left( \frac{C_{\rm s}}{2.5 \times 10^4 \km \s^{-1}} \right)
\erg \s^{-1} .
\label{eq:Wgcoccon}
\end{split}
\end{equation}
 Because the ram pressure at the stalled shock goes as $R^{-5/2}_{\rm R_s}$, and the sound speed also decreases somewhat, the rate of work varies with the shock radius as $W_{\rm g} \propto R^{-1}_{\rm s}$. For a shock radius of $R_{\rm s} = 200 \km$ the rate of work as given by equation (\ref{eq:Wgcoccon}) would be about half the value as given now in that equation. On the other hand, the value of $b$ might be somewhat larger. Still, the largest uncertainty is in the value of $\delta$.     

I comment on the two main assumptions that I used in deriving equation (\ref{eq:Wgcoccon}). (1) The gain region expands at the sound speed into the cocoon. The reason is that, as I discussed above, the pressure in the gain region is much larger than the pressure of the cocoon. The typical velocity by which a high pressure zone expands into a much lower pressure zone is the sound speed. In other words, I assume that the initial thermal and kinetic energy of the cocoon near the stalled shock is negligible. (2) The pressure inside the stalled shock does not drop despite the expansion along the cocoon. This assumption is based on the inequality $b \delta \ll 1$. Namely, the area through which the gas in the gain region expands out is small. Collapsing core material continues to fall on most of the other area of the stalled-shock sphere and therefore maintains the pressure there. Moreover, if the pressure in the gain region drops the stalled shock moves inward and this, as I indicated above, increases the work done according to equation (\ref{eq:Wgcoccon}) as $R^{-1}_{\rm s}$.  

The nucleon outflow rate in the two jets is $\dot M_{\rm 2j}/m_{\rm n}$, where $m_{\rm n}$ is the neutron mass. 
However, the outward flow of the material from the gain region will continue until the collapsing core material resumes its inflow onto the stalled shock in that region. 
By equation (\ref{eq:Tauhead}) the jet interaction with the collapsing core might last for about twice the duration of the jet-launching episode. Then, the ram pressure will return to its undisturbed value by about the time the collapsing core freely falls from that distance. This time might be a fraction of a second. From $r_{\rm h} = 3000 \km$ the free fall time is $\simeq 0.4 \s$.
Namely, more than 10 times the duration of the jet-launching episode. However, material from the sides of the jet's axis will partially fill the empty zone because the collapsing core at large distances still maintains a high thermal pressure. 
Overall, I take the outflow time of the gain region material to be several times the jet's activity phase $\tau_{\rm g,out} = {\rm several} \times \tau_{\rm j}$. 
  In section \ref{subsec:timescale} I return to discuss this timescale. 
The energy that the work of the gain region adds to the cocoon per nucleon in the two jets is then 
\begin{equation}
\begin{split}
& \Delta E_{\rm n_j,g} \approx
\frac{\tau_{\rm g,out}}{\tau_{\rm j}}
\frac{W_{\rm g}}{\dot M_{\rm 2j}/m_{\rm n}}  
\approx 50 
\left( \frac{\tau_{\rm g,out}}{10 \tau_{\rm j}} \right)
\\ \times
& \left( \frac{W_{\rm g}}{10^{51} \erg \s^{-1}} \right) 
\left( \frac{\dot M_{\rm 2j}}{0.1 M_\odot \s^{-1}} \right)^{-1}
\MeV. 
\label{eq:WgNucleon}
\end{split}
\end{equation}

The value of the energy per nucleon in the jets that the expanding gain region adds to the energy of the cocoon, i.e., the outflow that the jets induce, according to equation (\ref{eq:WgNucleon}) is highly uncertain and very crude. Nonetheless, equation (\ref{eq:WgNucleon}) does suggest that the outflow that the jets induce allows the high-pressure material in the gain region to expand and add a significant amount of energy relative to that of the jets, $E_{\rm n_j,k} = 57 \MeV$ in the present setting. 

\subsection{Relation to the standing accretion shock instability (SASI)} 
\label{subsec:SASI}

In the SASI the stalled shock surface oscillates with a large departure from spherical symmetry (e.g., \citealt{Blondinetal2003, Ohnishietal2006}). Because the ram pressure of the collapsing core varies as $\propto R^{-5/2}$ (e.g., \citealt{BlondinMezzacappa2006}), the ram pressure on a small protrusion of the stalled shock surface into the collapsing core (towards larger radius) is lower than on the rest of the stalled shock. This protrusion then grows. Simulations find that the $l=1$ and $l=2$ modes dominate the SASI (e.g., \citealt{Ohnishietal2006, BlondinMezzacappa2006, Hankeetal2013}). 
\cite{ Blondinetal2003} worked out the physics of SASI and write that if neutrino heating supports the stalled shock for a sufficiently long period of time the SASI might initiate the explosion. \cite{ Blondinetal2003} further discussed the possible coupling between rotation, magnetic fields, and SASI, as later numerical simulations show (e.g., \citealt{Summaetal2018}).

The effect by which the gain region does work on the cocoon, i.e., the shocked material of the jets and of the collapsing core around the jets (section \ref{subsec:AcceleratingCocoon}) is similar to the way the SASI facilitates explosion in the frame of the delayed neutrino explosion mechanism. 
The differences are that in the jittering jets explosion mechanism the perturbations that the jets cause are non-linear to start with (see Fig. \ref{Fig:Schematic} and \ref{Fig:Schematic2}), and that they are on smaller scales.  
Consider that the diameter of the cross section of the cocoon on stalled shock surface (Fig. \ref{Fig:Schematic2}) is $D \simeq (8 b \delta)^{1/2} R_{\rm s}$, where $b$ and $\delta$ are given by equations (\ref{eq:Acocoon}) and (\ref{eq:OmegaJets}), respectively. This corresponds to a mode of order 
\begin{equation}
l_{\rm j} \simeq \frac{\pi R_{\rm s}}{D} \simeq 6  
 \left( \frac{b}{3} \right)^{-1/2} 
\left( \frac{\delta}{0.01} \right)^{-1/2} .
\label{eq:Ljets}
\end{equation}
 
It is quite likely that the next jet-launching episode starts while the outflow from the gain region along the jets of the previous jet-launching episode has not ended yet. Namely, for a short time the perturbations by the jets occur at four places simultaneously. 

The conclusion is that the jets might excite SASI-like oscillations with $l_{\rm j} \simeq 6$ in addition to the $l=1$ and $l=2$ of the SASI itself. 

 In a recent study \cite{Vartanyanetal2022} examine the turbulence that develops below the stalled shock as a result of pre-collapse perturbations in the core. They find that the power spectra of tangential velocities in the turbulence zone is relatively flat up to spherical harmonic index $l \la 4$, and from there on a Kolgomorov-like spectrum forms. 
In the range $4 \la l \la 10$ the modes have energies $\simeq 0.2-1$ (depending on the 3D progenitor model) times the energy of the $ l \la 4$ modes. However, the seed perturbations in those simulations are small (in the linear regime). Here I suggest that the non-linear perturbations that the jets introduce might excite the $l \simeq 6$ SASI modes. My suggestion requires of course to be confirmed by 3D simulations.

\section{On the assumptions} 
\label{sec:Assumptions}

I here discuss the challenges and difficulties that the proposed boosting process should overcome before it stands on solid ground. 

\subsection{Motivation} 
\label{subsec:Motivation}

Before discussing the challenges of the jittering jets explosion mechanism I comment that there is a place to discuss the jittering jets explosion mechanism because of the imprints of jets in many supernova remnants (e.g., \citealt{Bearetal2017, GrichenerSoker2017, YuFang2018, Luetal2021, Soker2022a}), and because of the difficulties of the delayed neutrino mechanism. Among several difficulties  (e.g., \citealt{Kushnir2015, Papishetal2015}), the two main disadvantages of the delayed neutrino mechanism are that the maximum explosion energy that can be explained is $\simeq 2 \times 10^{51} \erg$ (e.g. \citealt{Fryeretal2012, Sukhboldetal2016}), and that in some cases explosion does not occur at all (e.g., \citealt{Burrowsetal2020}), although in many other cases explosion does take place in simulations (e.g. \citealt{Bolligetal2021, BurrowsVartanyan2021}).

\subsection{The available explosion energy} 
\label{subsec:Energy}

According to the jittering jets explosion mechanism the pre-collapse inner one or two convective zones launch the jets as they are accreted onto the proto-NS (newly born NS) through an intermittent accretion disk/belt at a radius of $\simeq 20-40 \km$. At the very beginning the proto-NS radius is $\simeq 50 \km$, but it shrinks to these smaller values by the time the jittering jets are launched. In some cases the mass in these convective zones is only $\Delta m_{\rm conv}=0.03 M_\odot$ \citep{ShishkinSoker2022}, and therefore do not supply by themselves enough mass to the jets to explain typical CCSN energies. However, in many other cases the mass in the pre-collapse convective zones is larger, and so the sample of stellar models cover the typical range of CCSN explosion energies \citep{ShishkinSoker2022}.  
For a proto-NS mass of $1.4 M_\odot$ and a radius of $R_{\rm PNS}=30 \km$ the escape velocity is $v_{\rm es}=1.1 \times 10^5 \km \s^{-1}$. If the jets carry a fraction of $f_{\rm 2j}=0.1$ of the accreted mass, then for the above values the energy that they carry will only be $\simeq 0.1 \Delta m_{\rm conv} v^2_{\rm es}/2 \simeq 3.6 \times 10^{50}$. This can explain by itself low-energy CCSNe. 
However, the available energy is larger. The reason is that the convective stochastic motion in the accreted layers serves as the seeds of the perturbations. Instabilities inside the stalled shock, like the spiral-SASI, amplify these seed perturbations and increase the amplitude of the specific angular momentum fluctuations, and, live longer to increase the amount of mass accreted through the intermittent disk/belt (see discussion in \citealt{ShishkinSoker2022}). 

Moreover, according to equation (\ref{eq:WgNucleon}) neutrino heating can double the original jets energy. Substituting typical values and a proto-NS mass of $1.4 M_\odot$ gives the total available energy according to the present study 
\begin{equation}
\begin{split}
 E_{\rm jets-total} & >  2 f_{\rm 2j} \Delta m_{\rm conv} v^2_{\rm es} /2 
=7 \times 10^{50} 
\\   & \times 
\left( \frac{\Delta m_{\rm conv}}{0.03 M_\odot} \right) 
\left( \frac{R_{\rm PNS}}{30 \km }\right)^{-1} 
\left( \frac{f_{\rm 2j}}{0.1} \right) \erg 
\label{eq:Ejetstotal}
\end{split}
\end{equation}
This is compatible with observed energies of  CCSNe as in most cases $\Delta m_{\rm conv} > 0.03 M_\odot$ \citep{ShishkinSoker2022}. 

\subsection{The timescale of the boosting process} 
\label{subsec:timescale}
\subsubsection{Comments on equation (\ref{eq:WgNucleon})}  
\label{subsubsec:Equation19}
In section \ref{subsec:AcceleratingCocoon} I estimated the outflow timescale
of the gain region material to be several times the jet’s
activity phase, $\tau_{\rm g,out} = {\rm several} \times \tau_{\rm j}$. 
As stated, this estimate should be confirmed with 2D or 3D numerical simulations. 
Here I further describe the process and the uncertainty in this time scale. 

After the jet activity ceases, the material in the gain region closes the funnel that each jet opened in the post-stalled shock zone in about a sound crossing time, 
$\Delta t_{\rm funnel,g} \simeq D/C_{\rm s} (R_{\rm s}) \simeq 5 \times 10^{-4} \s$ for a jet with a half opening angle of $8^\circ$ and sound speed as in equation (\ref{eq:Cs}). This very short time scale works for the proposed mechanism because it implies that practically the gain region starts to impart force on the cocoon (Fig. \ref{Fig:Schematic2}) immediately. 

The cocoon itself closes the funnel that the jet has opened outside the gain region (above the stalled shock). Because the cocoon is the post-shock media after it expanded, its temperature is lower than the immediately post-shock temperatures of the shocked collapsing core and of the jet near its head. Nonetheless, its sound speed is expected to be very large, definitely much larger than $1 \%$ of the jet velocity, i.e., $C_{\rm s} (\rm cocoon) > 1000 \km \s^{-1}$. It will close the funnel at $r_{\rm h} = 3000 \km$ in a time scale of $\Delta t_{\rm funnel,3000} < 0.4 \s$ for the same parameters that I use throughout the paper (see equation \ref{eq:Rhead}). As this is about the time I estimated, $\tau_{\rm g,out} = 10 \tau_{\rm j} \simeq 0.3 \s$ (see equation \ref{eq:WgNucleon}), I conclude that the cocoon also closes the funnel at any radius $r$ before the outflow that the gain region accelerates reaches that radius. 

I emphasize that although the cocoon closes the funnel, unlike the collapsing core, the cocoon does not fall towards the center at a high speed. As well, due to its high thermal pressure as a post-shock gas, its density is lower than that of the collapsing core. In other words, the jets pushed the gas to the sides as they expand outward. This will allow the high-pressure gas in the gain region to expand and further accelerate the cocoon to high velocities.   

To summarize this section I emphasize again the implication of equation (\ref{eq:WgNucleon}) and the discussion above. What I have shown is that the mechanism by which the delayed neutrino explodes stars, when it manages to work, is more efficient when we consider jets that locally revive the stalled shock. If this mechanism does not work here, the delayed neutrino mechanism will not work either.
What I further argue is that in many cases where the delayed neutrino mechanism does not work, the jittering jets explosion mechanism can work, and further be boosted by neutrino heating. As well, the jittering jets mechanism can account for explosion energies much above $2 \times 10^{51} \erg$. 

\subsubsection{On the possibility of prolonged jet activity} 
\label{subsubsec:prolonged}
The jittering jets mechanism was developed to explode starts within about a second from core bounce, similar to the expectations from the delayed neutrino mechanism in the past. Some recent studies suggest that the explosion activity can last for over five seconds (e.g., \citealt{Bolligetal2021}), while other studies, however, argue that the explosion should take place within less than a second (e.g., \citealt{Saitoetal2022}). The accretion rate in a fraction of the first second is indeed close to $1 M_\odot \s^{-1}$, as I scale quantities in this study. The accretion rate in the simulation of \citealt{Bolligetal2021} from $t=1 \s$ to $t=6 \sec$ is about $0.01 M_\odot \s^{-1}$. The total accreted mass might be $\simeq 0.05 M_\odot$ in this time period. The proto-NS radius is smaller at these late times $R_{\rm PNs} \simeq 15-20 \km$. Substituting this accreted mass and the smaller proto-NS radius in equation (\ref{eq:Ejetstotal}), I derive an energy that might be comparable to or larger than the energy that the jets carry in the first second. Namely, the jets can supply more energy if the process of accretion continues.

The prolonged jet activity has the advantage that at later times core material from further out is accreted, and those zones further out might in some cases posses  strong pre-collapse convection that seeds the instabilities that feed the stochastic angular momentum of the accreted gas (e.g., \citealt{ShishkinSoker2022}).

\subsection{The launching of jets} 
\label{subsec:Launching}
  
The processes that most severely needs confirmation by 3D simulations is the launching of the jittering jets by the intermittent accretion disk/belt that the accreted mass with stochastic angular momentum forms. 
These simulations must include magnetic fields (e.g., \citealt{Soker2018arXiv, Soker2019RAA, Soker2020RAA}), and be of very high resolution. For that, I do not expect simulations at present to be able to launch jets. 

Critical to the launching process is the formation of two opposite funnels near the proton-NS. The accretion disk/belt launches the two opposite jets through these funnels (e.g., \citealt{SchreierSoker2016}). Like the disk, the funnels are intermittent and change their direction in a stochastic manner, giving rise to jittering jets. Magnetic fields can form such funnels, as well as the stochastic angular momentum of the accreted gas. But even when the angular momentum of the accreted gas opens funnels along the angular momentum axis, magnetic fields are crucial to channel accretion energy to the collimated outflow. 

As I mentioned in section \ref{sec:intro}, the simulations by \cite{Kaazetal2022} suggest that such a process is possible. 
\cite{Kaazetal2022} simulate cases of a BH that accretes mass as it moves through a magnetized homogeneous medium, i.e., the Bondi-Hoyle-Lyttleton accretion flow. The initial direction of the magnetic field lines is perpendicular to the direction of BH motion through the medium. 
They obtain jets more or less along the initial direction of the magnetic field lines, despite that there is no initial angular momentum in the flow. However, for jets to be launched the magnetic fields have to be sufficiently strong. 

I therefore conclude that the fact that present simulations do not obtain jittering jets cannot be used to rule out such jets, simply because these simulations do not have yet all ingredients that might lead to such jets. 

\section{Summary} 
\label{sec:Summary}

I conducted a study to crudely estimate the boosting of jittering jets by neutrino heating. In the jittering jets explosion mechanism where there are several or more jet-launching episodes, the jets of one jet-launching episode are active for a relatively shot time and do not carry enough energy to explode the core. It is the additive effect of several and more jet-launching episodes that eventually explodes the star. 

 I chose a set of parameters for the ambient gas and jets (section \ref{sec:RelevantParameters}) to perform the calculations. 
I then studied two processes by which neutrino heating can increase the energy that the jets deposit to the collapsing core material. 

In the direct heating process (section \ref{subsec:HeatingCocoon}) the neutrinos that the cooling proto-NS emits directly heat the cocoon (yellow-hatched region in Fig. \ref{Fig:Schematic}). I find that this direct heating process adds less than about ten percent of the kinetic energy that the jets carry, $\Delta E_{\rm n_j, \nu}  < 0.1 E_{\rm n_j,k}$, where these energies are defined per nucleon in the jets (equation \ref{eq:DeltaEn}). 

In the second process (section \ref{subsec:AcceleratingCocoon}) the material from the gain region flows out into the cocoon and does work on it. The material in the gain region maintains its high pressure by neutrino heating. I estimate the work that the gain region does on the cocoon to be crudely equal to the initial energy that the jets carry, $\Delta E_{\rm n_j,g} \approx E_{\rm n_j,k} $ (equation \ref{eq:WgNucleon}). 

In section \ref{subsec:SASI} I raised the possibility that the jets excite SASI-like oscillations but with higher orders of $l \simeq 6$ (equation \ref{eq:Ljets}) with respect to the regular SASI models that have mainly $l=1$ or $l=2$. These modes exist alongside the regular SASI modes.  

\cite{ Blondinetal2003} mentioned the possible coupling between neutrino heating, the SASI, rotation, and magnetic fields, in exploding CCSNe. The spiral modes of SASI amplify initial perturbations to supply the stochastic angular momentum to  the accreted gas that launches the jittering jets (e.g., \citealt{ShishkinSoker2021}), and strong magnetic fields are involved in launching the jets (sections \ref{sec:intro}  and \ref{subsec:Launching}). 
In the present study I strengthen the claim for a mutual influence between jittering jets and neutrino heating. I therefore extend the statement of \cite{Blondinetal2003}  to include also jets, and argue that the coupling between spiral-SASI, stochastic rotation, possible ordered rotation, magnetic fields, and \textit{(jittering) jets}, lead to the explosion of CCSNe. 

 In section \ref{sec:Assumptions} I discussed some of the assumptions of the present study. Although the launching of jittering jets is impossible to include in present numerical studies  (section \ref{subsec:Launching}), I encourage simulations to insert jittering jets near the stalled shock and study the effects of neutrino heating on these jets and the outflow they induce. 

\section*{Acknowledgments}

I thank Dima Shishkin for helpful comments, and an anonymous referee for detailed and very useful comments. This research was supported by a grant from the Israel Science Foundation (769/20).

\textbf{Data availability}
The data underlying this article will be shared on reasonable request to the corresponding author.  


\end{document}